\documentstyle[12pt,epsfig]{article}
\newcommand{\alpi}{\it O(\alpha/\pi)}
\def\beq{\begin{equation}}
\def\eeq{\end{equation}}
\def\amu{a_\mu}
\def\amuSM{a_\mu^{\rm SM}}
\def\amususy{a_\mu^{\rm SUSY}}

\def\tanbeta{{\rm tan}\beta}
\def\gmin2{(g-2)_\mu}

\def\m3half{m_{3/2}}

\def\chargino1{\tilde \chi^\pm_1}
\def\neut1{\tilde \chi^0_1}
\def\smallchargino{m_{ {\tilde \chi}_1^{\pm}}}
\def\largechargino{m_{ {\tilde \chi}_2^{\pm}}}
\def\muonsneutrino{m_{{\tilde \nu}_\mu}}
\def\smallneut{m_{{\tilde{\chi}_1^0}}}

\def\staumass{m_{{\tilde{\tau}_1}}}
\def\stau{{\tilde \tau}_1}
\def\r2{\sqrt 2}
\def\cb{\cos\beta}
\def\sb{\sin\beta}
\catcode`\@=11 % This allows one to modify PLAIN macros.

\def\lsim{\mathrel{\mathpalette\@versim<}}
\def\gsim{\mathrel{\mathpalette\@versim>}}
\def\@versim#1#2{\vcenter{\offinterlineskip
    \ialign{$\m@th#1\hfil##\hfil$\crcr#2\crcr\sim\crcr } }}

\def\PRL{Phys. Rev. Lett.}

\def\NP{Nucl. Phys.}

\begin{document}
\begin{flushright}
{TIFR/TH/00-29 \\
hep-ph/0006049}
\end{flushright}
%-----------------------------------
%\documentstyle[preprint,aps]{revtex}
\begin{center}
{\Large\bf 
Constraining Anomaly Mediated Supersymmetry Breaking Framework 
via Ongoing Muon $ g-2 $ Experiment at Brookhaven
\\}
\vglue 0.5cm
{\bf Utpal Chattopadhyay$^a$\footnote{utpal@theory.tifr.res.in, 
Dilip.Ghosh@cern.ch, sourov@theory.tifr.res.in}, 
Dilip Kumar Ghosh$^b$\footnote{On leave of absence from Department of 
Theoretical Physics, TIFR, Mumbai, India.},
and Sourov Roy$^a$}

\vglue 0.2cm
{\em 
$^a$Department of Theoretical Physics\\
Tata Institute of Fundamental Research\\
Homi Bhabha Road, Mumbai 400 005, India\\
\vspace{0.4cm}
$^b$CERN, Theory Division, CH-1211 Geneva 23, Switzerland \\
}
\end{center}

\vskip 15pt

\begin{center}
{\bf                             Abstract
}
\end{center}

\begin{quotation}
\noindent
The ongoing high precision E821 Brookhaven National Laboratory 
experiment on muon $ g-2 $ is promising to probe a theory involving 
supersymmetry. We have studied the constraints on the minimal Anomaly 
Mediated Supersymmetry Breaking (AMSB) model using the current data 
of muon $ g-2 $ from Brookhaven. A scenario of seeing no deviation 
from the Standard Model is also considered, within a $2\sigma$ limit 
of the combined error from the Standard Model result and the Brookhaven 
predicted uncertainty level. The resulting constraint is found to be 
complementary to what one obtains from $b \rightarrow s+ \gamma$ bounds 
within the AMSB scenario, since only a definite sign of $\mu$ is 
effectively probed via $b \rightarrow s+ \gamma$. A few relevant generic 
features of the model are also described for disallowed regions of 
parameter space.

\vskip 15pt


\end{quotation}

\setcounter{footnote}{0}   

\section{Introduction}
   The search for supersymmetry (SUSY) in high energy physics relies both on 
high energy colliders as well as on experiments based on 
perturbative corrections to various experimentally measurable 
quantities. Traditionally, the measurement of an electron's
anomalous magnetic moment has been highly effective in verifying the 
prediction of quantum electrodynamics (QED) to a very high 
order. Probing Beyond the Standard Model physics with supersymmetry  
is seen to be possible with a precision $(g-2)$ measurement 
of the muon. The ongoing muon $(g-2)$ measurement E821\cite{recentBNL} at 
Brookhaven National Laboratory (BNL), designed to verify the results of 
Standard Model (SM) electroweak corrections, has already provided a more accurate 
result than the previous CERN experiment\cite{CERNg},
by a factor of 2 or so. With improved design and state of the art 
technology, it is expected that within a few years 
from now, the accuracy of the BNL result will be increased by a 
factor of 20, or even more, compared to the same of the previous CERN 
measurement.

Supersymmetric electroweak corrections to muon $(g-2)$
can be as large as the SM electroweak correction, and this fact has been
seen in a number of references ranging from the minimal supersymmetric
standard model (MSSM)\cite{tipnCPg,gMSSM},
Supergravity based 
models\cite{ucpn,gSUGRA}, and Gauge Mediated Supersymmetry Breaking 
scenarios\cite{ggmsb}.
In the recent past, considerable interest has been seen 
in a different type of SUSY breaking mechanism, other than Supergravity
\cite{pnsugra} and Gauge Mediated SUSY Breaking\cite{gmsbref}, which at its 
dominating scenario, is generically known as Anomaly Mediated Supersymmetry 
Breaking (AMSB) [\ref{randall}--\ref{bagger}].
This effect originates from the existence of a super Weyl anomaly\cite{randall} 
while considering SUSY breaking. 
As we will discuss latter, the problem of the resulting
tachyonic sleptons which arises within the AMSB sparticle spectrum, 
is avoided in the 
{\it minimal} definition [\ref{feng_moroi_g-2}--\ref{kribs}]
of the model, via adding a common scalar mass $m_0$ 
with all the scalars of the theory, at a given scale.
%(changed)As we will discuss latter, the resulting
%tachyonic sleptons within AMSB sparticle spectrum is avoided in the 
%{\it minimal} definition[\ref{feng_moroi_g-2}--\ref{kribs}]
%of the model, by adding a common scalar mass $m_0$ with all the scalars 
%of the theory at a given scale.

Large SUSY contributions 
$\amususy \equiv {1 \over 2}(g-2)_\mu^{\rm SUSY}$ in the minimal AMSB 
framework have already been seen in 
Ref.~\cite{feng_moroi_g-2}, in which the authors discussed in 
detail a broad range of interesting phenomenological 
implications involving colliders, as well as 
various low energy signatures within the 
model, in addition to
showing that the minimal model may remain {\it natural} even for
a superheavy $m_0$.
In this work, we will analyze the constraint coming from $\amususy$
with regard to the high precision Brookhaven 
experiment, within the minimal AMSB framework. 
Considering $a_\mu^{\rm expt} - \amuSM = 
\amususy$ and taking into account the associated error limits in 
quadratures, we will constrain the SUSY parameter space of the 
model.

Our work will be organized as follows. In Sec. 2, we will discuss the 
SM result for $\amu$ for its different parts of contributions. 
We will see the existence of a large error associated
with the lowest order hadronic vacuum polarization contribution 
and sources of its possible improvement in the near future, via low energy 
$e^+e^- \rightarrow hadron$ data from various experiments.
We will describe the AMSB framework
and the necessity of defining a {\it minimal} scenario in Sec. 3.
In Sec. 4, we will use Ref.~\cite{ucpn} for the result of $\amususy$,
where the analysis was performed in the minimal 
supergravity (mSUGRA) model to see the constraint from the high precision
BNL experiment. 
We will analyze the constraint from $\amususy$ 
on the parameter space of the
minimal AMSB model, due to the present SM and
experimental results of $\amu$.
We will also investigate the potential for constraining the minimal AMSB
model using the predicted level of the uncertainty of the E821 BNL experiment.
We will do so in the minimal {\it no-deviation} scenario, which here means seeing no
disagreement from the SM result within the experimental and the theoretical
uncertainties, once the measurement is complete at the desired level of accuracy.
We will also quote the result of the $b \rightarrow s +\gamma$ constraint for
both signs of $\mu$ and the positive role of an $\amususy$ analysis 
in this regard.
In Sec. 5 , 
we will comment on the disallowed regions which appear due 
to reasons other than $\amususy$ limits. 
Primarily, we will examine the disallowed regions of parameters space 
due to the combined effect of 
the scale invariant part of the scalar mass relations of 
sleptons with gauge and Yukawa couplings within the minimal AMSB model and 
the nature of the associated renormalization group evolutions. 
We will also see the effect of SUSY-QCD corrections to the bottom-quark 
mass on the minimal AMSB spectra, a large $\tanbeta$ effect, which also
has specific features within AMSB models.
These regions,
when combined with $\amususy$ eliminated parameter space,
provide simpler and definite predictions for the lower bounds of the masses of relevant 
the supersymmetric particles, as well as of the input parameters of the model.
  
\section{Standard Model result $\amuSM$ and sources of uncertainty}

\noindent
The Standard Model result for $\amu$ is
\beq
{1 \over 2}(g-2)_\mu^{\rm SM}= \amuSM=11659162.8(6.5) 
\times 10^{-10} \equiv (11659162.8 \pm 6.5) \times 10^{-10}  
\eeq
In contrast, the latest data from the ongoing E821 
BNL experiment\cite{recentBNL} amounts to:
\beq
a_\mu^{\rm expt} = 11659210 (46) \times 10^{-10}  
\eeq
The uncertainty amount is expected to be reduced to 
$ \delta \amu^{\rm BNL} \lsim 4 \times 10^{-10}$. 

\begin{table}[h]
\begin{center}
\caption{Contributions to $\amuSM$ (in units of $10^{-10}$)}
\begin{tabular}{|l|l|}
\hline
\hline
Nature of Contribution & Value \\
\hline
\hline
Q.E.D. to $\alpi^5$ \cite{qedrefs} ($a_\mu^{\rm QED}$)  & 11658470.6(0.3) \\
\hline
\hline
Hadronic vac. polarization to $\alpi^2$ ($a_\mu^{had1}$)\cite{Davier98} &692.4(6.2) \\
Hadronic vac. polarization to $\alpi^3$\cite{krause} &$-10.1(0.6)$ \\
Light by light hadronic amplitude\cite{hayakwa} & $-7.92(1.54)$ \\
\hline
\hline
Total hadronic ($a_\mu^{\rm hadronic}$)  & 677.1(6.5) \\
\hline
\hline
Total electro-weak up to 2-loops ($a_\mu^{\rm EW}$)   & 15.1(0.4) \\
\hline
\hline
$\amuSM$ & 11659160.1(6.5) \\ %sep24
\hline
\end{tabular}
\end{center}
\end{table}

\noindent
The SM result is broken up into
\beq
\amuSM = a_\mu^{\rm QED} + a_\mu^{\rm hadronic} + a_\mu^{\rm EW} .
\eeq
Here $a_\mu^{\rm QED}$ is the pure QED contribution computed up to 
five loops in electromagnetic coupling. The quantity 
$a_\mu^{\rm hadronic}$ refers to the
total hadronic contribution including the lowest order and the next to lowest 
order hadronic vacuum polarizations and the light-by-light hadronic 
contribution. The electroweak part $a_\mu^{\rm EW}$ is the SM
electroweak contribution up to two loops. The amounts from the 
individual parts with corresponding references are summarized in Table~(1).

Within $a_\mu^{\rm hadronic}$, the contribution $a_\mu^{had1}$ arising 
from the $\alpha^2$ level of hadronic vacuum 
polarization diagram is the least accurate quantity, and 
its uncertainty is almost the same as the overall error of $\amuSM$. 
Hence an accurate determination of
$a_\mu^{had1}$ will be increasingly important for compatibility with the
high-precision measurement at BNL, which has an expected level of 
uncertainty of $\delta \amu^{\rm BNL} \lsim 4 \times 10^{-10}$. 
However, the present uncertainty
in $\amuSM$ is quite small compared to the experimental uncertainty level, 
a situation which is going to be changed within a few years. 
The largest hadronic contribution $\amu^{had1}$
is obtained from the total Born cross section (lowest order in QED)
for hadron productions in $e^+e^-$ annihilation, a result found via
dispersion theory and optical theorem\cite{OPTtheo}. Accurate
low energy $e^+e^- \rightarrow hadrons $ data hence become 
necessary to lower the uncertainty level.
Measurements of low energy hadron production cross sections 
in BES, CMD-II and DA$\Phi$NE\cite{exphad}, 
will significantly improve the result
of $\amuSM$. 
In the recent past, the authors of Ref.~\cite{Davier98} used ALEPH data 
from hadronic $\tau$ decay, and QCD sum-rule techniques
to evaluate $a_\mu^{had1}$ and this improved 
the hadronic error estimate very significantly. Our analysis
uses this result.
Further prospects for improving the result of $a_\mu^{had1}$, as well 
as a critical evaluation of the above estimate may be seen 
in Ref.~\cite{jegerlehner}.

\section{Anomaly Mediated Supersymmetry Breaking in the Minimal Scenario}
In a supersymmetric theory, additionally, 
soft SUSY breaking parameters are also contributed via 
Super-Weyl anomaly, which is a generic feature if SUSY is 
broken in a supergravity framework. In anomaly mediated supersymmetry
breaking scenario, the Super-Weyl anomaly contributions dominate,
because the SUSY breaking sector and the visible sector reside 
in different parallel 3-branes\cite{randall}
in a string theoretical perspective, and there are no tree-level
couplings between the two branes. 
The form of soft parameters 
thus generated, are renormalization group (RG) invariant, and at any 
desired scale, the soft parameters are determined by the appropriate 
gauge and Yukawa couplings for the same scale. 
This is particularly 
interesting to avoid a SUSY flavor problem, because
the scale invariance of soft parameters which is provided
by special RG trajectories, eliminates the effect of any
flavor violating unknown physics possibly existing at a higher scale.

In spite of many desirable features of anomaly mediation, within 
the framework of the MSSM, one finds 
that sleptons become tachyonic.
In the {\it minimal} AMSB model such tachyonic sleptons are avoided
by introducing an additional common mass parameter $m_0$ for all 
the scalars of the theory.
But, this obviously violates the scale invariance of the model, 
whereas preserving the same would be desirable in regard 
to the flavor changing neutral current (FCNC) constraint. 
Tachyonic sleptons are avoided differently in non-minimal AMSB 
models\cite{nonminimalref}, which have appropriate scale invariant scalar 
mass combinations within RG evolutions, but these are outside the scope of 
our present work. However, via the existence of focus point
\cite{fengfocus} of the renormalization group equation
(RGE) of $m_{H_u}^2$, the 
minimal model allows the possibility of multi-TeV squarks and 
sleptons without increasing the 
fine-tuning measure\cite{feng_moroi_g-2}, and this is an 
important feature to address many phenomenological issues.

Scalar masses are hence determined via renormalization group equations
of the MSSM starting from their respective values at the unification scale 
($M_G \sim 1.5$ to $2.0\times10^{16}$ GeV) and leading up to the electroweak scale
$M_Z$, the scale for mass of the Z-boson.
However, for the first two generations of scalars,  because of the 
negligible first two generation Yukawa couplings, the effect translates 
to having simply an overall additive constant $m_0$ at $M_G$, which would
have minimal changes due to RG evolutions.
 
In this analysis, the evolutions of scalar masses,
gauge and Yukawa couplings are computed at two loop level of
RGE\cite{martin94}, and trilinear couplings are evolved via one-loop 
level of the same.  
Unification of gauge couplings is incorporated with having
$\alpha_3(M_Z) \sim 0.118$.
For the Higgsino mixing, $\mu^2$ is computed radiatively via electroweak 
symmetry breaking condition at 
the complete one loop level of the effective 
potential\cite{oneloopeff}, while 
optimally choosing a renormalization scale  
$Q = \sqrt{(m_{{\tilde t}_1}m_{{\tilde t}_2})}$ 
for minimization. 
The analysis also includes supersymmetric QCD 
correction to bottom quark mass\cite{susybmass},
which is considerable for large $\tanbeta$ regions
also having its important features in AMSB scenarios.

With a high degree of predictivity the model is 
described by the following parameters:
the gravitino mass $m_{3/2}$, the common scalar 
mass parameter $m_0$, the ratio of Higgs vacuum expectation values 
$\tanbeta$, and $ sgn(\mu)$. 
Following Ref.~\cite{baer_many}, with having 
the same sign conventions for $\mu$ and $A$-parameters in this work, 
we see that the masses are given via the couplings as follows.

\noindent
{\it Gauginos:}

\begin{equation}
{\tilde m}_1= {33\over 5}{g_1^2\over 16\pi^2}m_{3/2}, \quad
{\tilde m}_2= {g_2^2\over 16\pi^2}m_{3/2}, \\
{\tilde m}_3= -3{g_3^2\over 16\pi^2}m_{3/2} 
\label{gauginoeqn}
\end{equation}

\noindent
{\it Higgs and Third generation scalars:}
\beq
{\tilde m}_i^2 = C_i {m_{3/2}^2 \over {(16\pi^2)}^2} + m_0^2
\label{thirdgeneqn}
\eeq
where $i \equiv (Q,U,D,L,E,H_u,H_d)$, with $C_i$'s being given as\\

$C_Q = -{11\over 50}g_1^4-{3\over 2}g_2^4+
8g_3^4+h_t\hat{\beta}_{h_t}+h_b\hat{\beta}_{h_b}$, 
$C_U = -{88\over 25}g_1^4+8g_3^4+
2h_t\hat{\beta}_{h_t}$ \\ 
$C_D = -{22\over 25}g_1^4+8g_3^4+
2h_b\hat{\beta}_{h_b}$, 
$C_L= -{99\over 50}g_1^4-{3\over 2}g_2^4+
h_\tau\hat{\beta}_{h_\tau}$, 
$C_E = -{198\over 25}g_1^4+
2h_\tau\hat{\beta}_{h_\tau}$ \\  
$C_{H_u} = -{99\over 50}g_1^4-{3\over 2}g_2^4+
3h_t\hat{\beta}_{h_t}$, and 
$C_{H_d} = -{99\over 50}g_1^4-{3\over 2}g_2^4+
3h_b\hat{\beta}_{h_b}+h_\tau\hat{\beta}_{h_\tau}$ \\  

\noindent
Here, Q and L are the superpartners of quark and lepton doublet fields, respectively.
The superpartners for singlet quark fields for up and down type are U and D, and 
the same for singlet lepton is E. 

\noindent
{\it Trilinear couplings:}

\beq
A_t = {\hat{\beta}_{h_t}\over h_t}{m_{3/2}\over 16\pi^2},\quad 
A_b = {\hat{\beta}_{h_b}\over h_b}{m_{3/2}\over 16\pi^2}, \quad {\rm and \quad}
A_\tau = {\hat{\beta}_{h_\tau}\over h_\tau}{m_{3/2}\over 16\pi^2},
\label{Aeqn}
\eeq

\noindent
where $\hat \beta$'s are defined by \\ \\
$\hat{\beta}_{h_t} = h_t\left( -{13\over 15}g_1^2-3g_2^2
-{16\over 3}g_3^2+6h_t^2+h_b^2\right)$, 
$\hat{\beta}_{h_b} = h_b\left( -{7\over 15}g_1^2-3g_2^2
-{16\over 3}g_3^2+h_t^2+6h_b^2+h_\tau^2\right)$, 
and $\hat{\beta}_{h_\tau} =h_\tau
\left( -{9\over 5}g_1^2-3g_2^2+3h_b^2+4h_\tau^2\right)$. \\

\noindent
The quantities for the first two generations can be obtained similarly 
by considering appropriate Yukawa couplings, which, however, are neglected 
in our analysis. We note that,
Eq.~(\ref{gauginoeqn}) and Eq.~(\ref{Aeqn}) are scale 
invariant.  
Hence, having no intrinsic
RG evolutions of their own, the masses and the trilinear couplings 
may be computed at any scale once the appropriate gauge and Yukawa 
couplings are known. 
However, because of the addition of the $m_0^2$ term, which rescues sleptons
from being tachyonic, Eq.~(\ref{thirdgeneqn}) is not scale invariant.
Here the scale for obtaining the mass values of 
Eq.~(\ref{thirdgeneqn}) is chosen as $M_G$. 
Thereafter, RG 
evolutions of soft scalar parameters and use of
the electroweak radiative breaking condition at the complete
one loop level produce the sparticle mass spectra. 
One of the important features of the minimal AMSB model is that
the resulting $SU(2)$ gaugino mass $\tilde m_2$ is quite smaller than
$\tilde m_1$ as well as $|\mu|$. 
Here, we have also incorporated the non-negligible next to leading 
order (NLO) corrections~\cite{wells} for gaugino masses. 
As a result, the lighter chargino
$\chargino1$ and lightest neutralino $\neut1$ are wino dominated; indeed
they are almost degenerate, with the latter becoming the lightest 
supersymmetric particle (LSP).
This has interesting phenomenological aspects, like what is seen most recently
in Ref.~\cite{ghosh}, where a definite signal in a linear $e^+e^-$ collider
could be predicted as a possible minimal AMSB signature. 
Compared to other SUSY scenarios where the lightest neutralino
has a distinctly smaller mass, here this similar mass value 
of $\smallchargino$
and $\smallneut$ effectively decreases $|\amususy|$, although 
a weakly contributing effect. 
Another striking result of the minimal AMSB model is the strong mass degeneracy
between left and right sleptons. Consequently, the third and the 
second generation {\it L-R} mixing angles become significantly 
larger, going up to the maximal limit for a large $\tanbeta$. We will see  
the strong effect of large smuon {\it L-R} mixing on the neutralino loop 
contributions of $\amususy$ in Sec. 4.
 
\section{Results} 
\begin{figure}[hbt]
\centerline{\epsfig{file=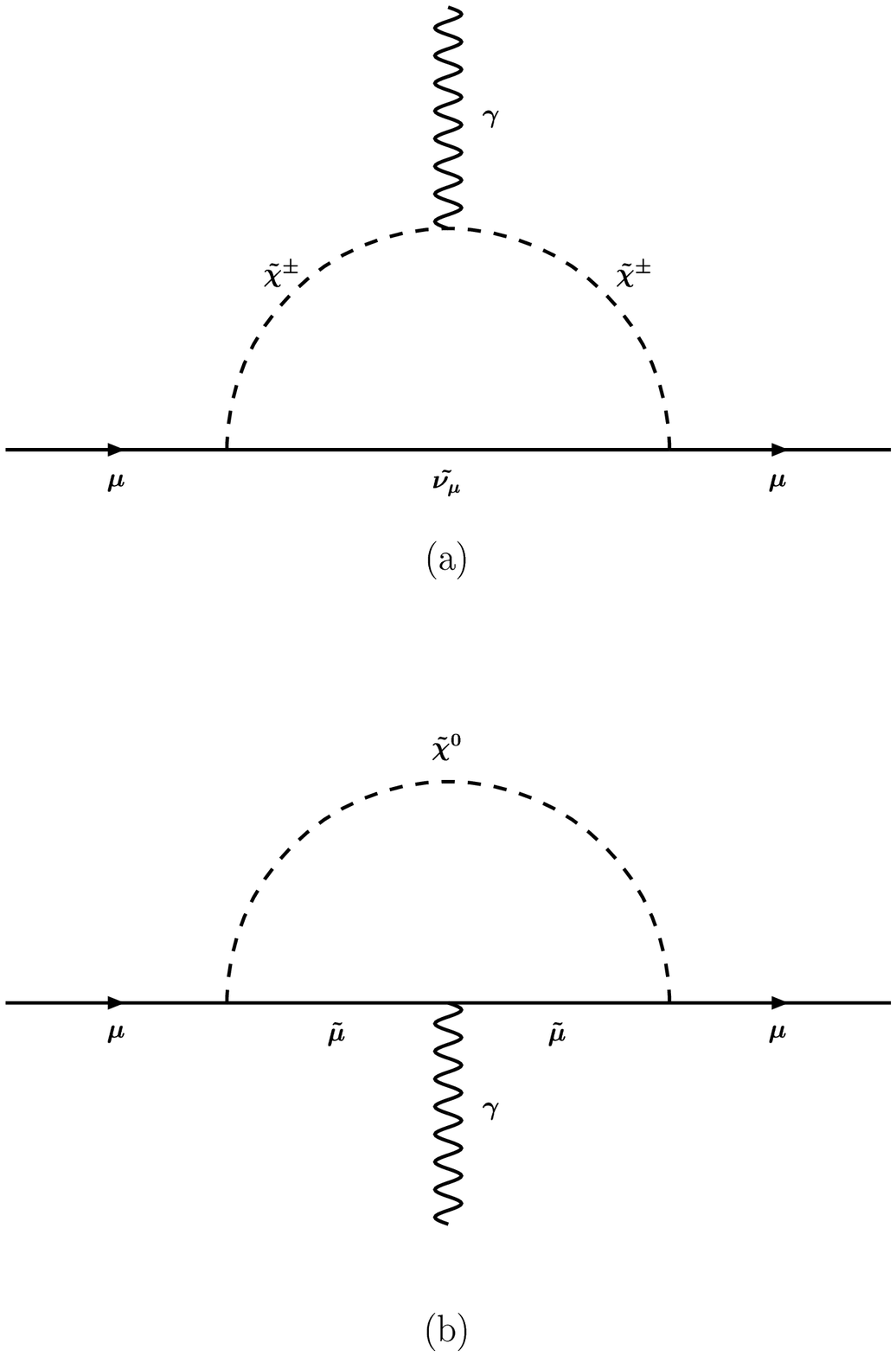,width=15cm}}
\vspace*{-1.5in}
\caption{Feynman diagrams contributing to ${a_\mu}^{\rm SUSY}$ 
(a) for the chargino-sneutrino loop, (b) for the neutralino-smuon loop.}
\label{feyn_diag}
\end{figure}

  The diagrams to compute the supersymmetric contributions to 
muon $(g-2)$, as shown in Figs.~(\ref{feyn_diag}a) and (\ref{feyn_diag}b), 
are divided into the chargino-sneutrino loop and the neutralino-smuon loop.
We only quote the chargino part of the result here, 
which dominates in $\amususy$.
The neutralino contribution may be seen in Refs.~\cite{tipnCPg,ucpn}
\footnote{The most general result of the SUSY 
electroweak contribution to muon $(g-2)$ in the MSSM, where {\it CP} violating 
phases are considered, can be seen in Ref.~\cite{tipnCPg}.}. 
Separating the chargino contributions into chirality diagonal and 
nondiagonal parts we have
\beq
{a_\mu^{\rm SUSY}}^{\chi^\pm}={a_\mu^{\rm SUSY}}^{\chi^\pm} ({\rm nondiag})+ 
{a_\mu^{\rm SUSY}}^{\chi^\pm}({\rm diag}),
\label{charginotot}
\eeq
where
\beq
{a_\mu^{\rm SUSY}}^{\chi^\pm} ({\rm nondiag}) 
=\frac{m^2_{\mu}\alpha_{em}}{4\r2\pi m_W\sin^2\theta_W \cb}
\sum_{i=1}^{2}\frac{1}{m_{\chi_i^\pm}}(U_{i2}V_{i1})
F\left(\frac{m^2_{\tilde{\nu_\mu}}}{m^2_{\chi_i^\pm}}\right)
\label{charginonondiag}
\eeq
and
\beq
{a_\mu^{\rm SUSY}}^{\chi^\pm}({\rm diag})
=\frac{m^2_{\mu}\alpha_{em}}{24\pi\sin^2\theta_W}
\sum_{i=1}^{2}\frac{1}{m^2_{\chi_i^\pm}}
\left(\frac{m^2_{\mu}}{2 m^2_W \cos^2\beta} U_{i2}^2+V_{i1}^2 \right)
G\left(\frac{m^2_{\tilde{\nu_\mu}}}{m^2_{\chi_i^\pm}}\right).
\label{charginodiag}
\eeq
Here, in general, {\it U} and {\it V} are unitary 
$2 \times 2$ matrices, which diagonalize
the chargino mass matrix $M_{\tilde \chi^\pm}$ as shown below, via a 
bi-unitary transformation, 
$U^* M_{\tilde \chi^\pm} V^{-1} ={\rm diag}(\smallchargino,\largechargino)$:
\beq
M_{\tilde \chi^\pm}=\left(\matrix{ {\tilde m_2} & \r2 m_W  \sb \cr
        \r2 m_W \cb & \mu} \right).
\eeq
With $M_{\tilde \chi^\pm}$ being real
{\it U} and {\it V} are orthogonal matrices. 
The functions $F(x)$ and $ G(x)$ arising from loop integrations 
are given by: $F(x)=(3x^2-4x+1-2x^2{\rm ln}x)/(x-1)^3$, and
$G(x)=(2x^3+3x^2-6x+1-6x^2 {\rm ln}x)/(x-1)^4$.

\begin{figure}[hbt]
\centerline{\epsfig{file=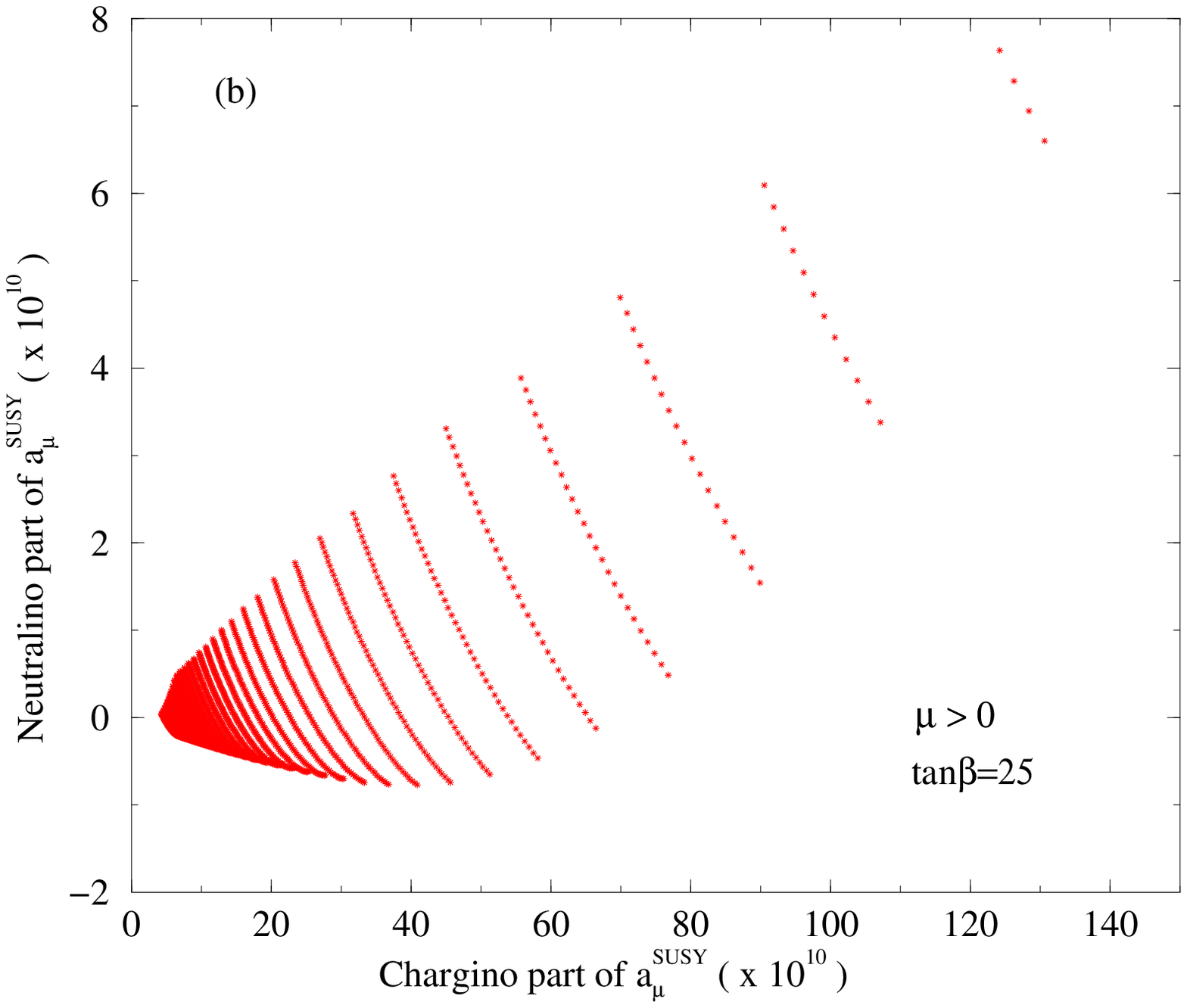,width=8cm}}
%\vspace*{-2.0in}
\caption{Relative contributions to muon ${a_\mu}^{\rm SUSY}$ for (a) 
$\mu < 0$ and (b) $\mu > 0$ from chargino-sneutrino and neutralino-smuon 
loops when $\tan \beta =25$, $m_0 \leq 1000$ GeV and $m_{3/2} 
\leq 100$ TeV.}
\label{relativefigp}
\end{figure}

Using the complete $\amususy$ result for numerical computations, 
Figs.~(\ref{relativefigp}a) and (\ref{relativefigp}b) show 
the dominance of chargino contributions over the neutralino parts.
Here, the two types of contributions are plotted along the axes 
for $\tanbeta$=25, when $\m3half$ and $m_0$ are varied over 
a broad range of values ($\m3half < 100$~TeV and $m_0 < 1$~TeV). 
It is indeed the chirality nondiagonal 
term involving the {\it lighter} chargino and sneutrino part which 
dominates over the other contributions in $\amususy$.
Because of the same reason, as explained further in a similar 
mSUGRA analysis of 
Ref.~\cite{ucpn}, there is a definite sign dependence 
between $\amususy$ and $\mu$, namely, $\amususy > 0$ for $\mu >0$, and 
$\amususy < 0$ for $\mu <0$, an important result for $\amususy$.
Thus, the lighter chargino mass ($\smallchargino$) has a significant role 
in $\amususy$. On the other hand, for a given 
$\smallchargino$ value, a heavier $\muonsneutrino$ decreases $|\amususy|$.
Because of the presence of 
Yukawa coupling ($\sim {1 \over \cos \beta} $) within the chirality
nondiagonal terms for both the chargino (see Eq.~\ref{charginonondiag})
and neutralino results, we see that $|\amususy|$ is almost 
proportional to $\tanbeta$. 

The special signature of AMSB due to an extremely large smuon {\it L-R} 
mixing angle, as mentioned before, affects the neutralino results to diminish strongly via partial 
cancellation between the terms.
The particular neutralino term (see Ref.~\cite{ucpn}), 
which involves smuon mixing, becomes almost comparable to the significantly
contributing chirality nondiagonal neutralino term associated with 
Yukawa coupling.  
Both terms depend on $\tanbeta$, as well as on the sign of $\mu$. 
A detail numerical investigation shows
that the two terms always come in opposite signs, giving
a large cancellation within the neutralino result. 
On a relative scale, we have seen that, for a 
naturalness~\cite{klucpn} favored region of SUSY spectra with a 
given $\tanbeta$, the ratio of neutralino to chargino 
contributions within $\amususy$ in minimal AMSB is 
typically smaller by 50\% or so, compared to the same 
within a similar natural mSUGRA spectra.  

\subsection{Constraints from present values of $\amu^{\rm expt}$ and $\amuSM$}
\begin{figure}[hbt]
\centerline{\epsfig{file= 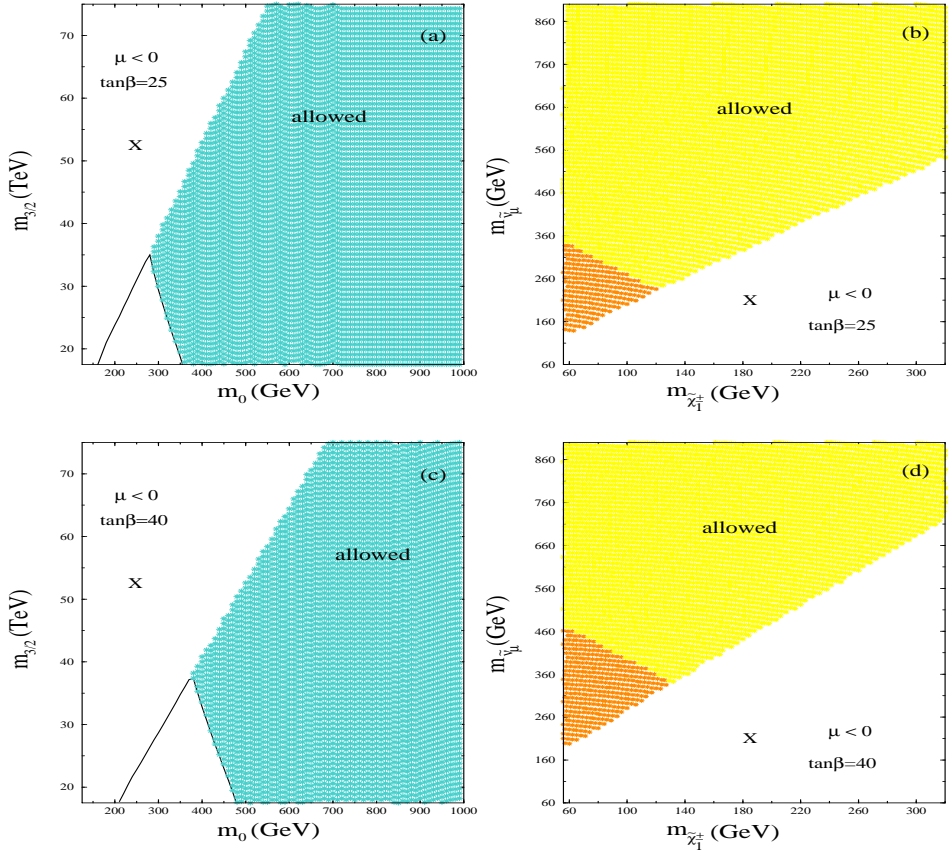,width=15cm}}
\vspace*{-3.5in}
\caption{Constraints in the $(m_0 - m_{3/2})$ and $(m_{\tilde \chi^\pm_1}- 
m_{\tilde \nu_\mu})$ planes from the present limits of $a_\mu^{\rm SUSY}$
for $\mu < 0$. Regions allowed 
by the $a_\mu^{\rm SUSY}$ constraint are labeled as {\it allowed}. Regions 
marked with {\it X} are generally disallowed zones for the SUSY spectra 
within minimal AMSB. Small white triangular regions below the {\it allowed}
areas in the left hand side 
figures are disallowed via  the $a_\mu^{\rm SUSY}$ limit. The same regions in the 
right hand side are darkly shaded.
}
\label{constraints_mpr}
\end{figure}

  Figures~(\ref{constraints_mpr}a) to (\ref{constraints_mpr}d)
show the constraints arising from $\amususy$ when the present 
experimental data from Brookhaven is compared with the Standard 
Model result. 
Here we consider the residual amount $\amu^{\rm expt} - \amuSM$ to limit
$\amususy$ within the 2$\sigma$ level of combined error estimates, added in 
quadratures ( $ -43.0 \times 10^{-10} <\amususy < 142.8 \times 10^{-10}$).
Considering the largest possible $|\amususy|$ within the model, we see 
that, essentially, a constraint exists only for $\mu<0$.
The regions excluded by $\amususy$ bounds when combined with the 
disallowed regions ( labeled by {\it X} ) characteristic of the 
minimal AMSB model itself, along with the experimental constraints on 
various sparticle masses, a value of $m_0$ below 275 GeV is completely 
eliminated for any value of $\m3half$ (see Fig.~(\ref{constraints_mpr}a)).  
The nature of the excluded region as marked by {\it X} principally
originates from sleptons turning into the LSP and then becoming 
tachyonic at the electroweak scale ( see Sec. 5 ).
The same for $\tanbeta=40$ as shown in Fig.~(\ref{constraints_mpr}c) 
amounts to $m_0 \sim 375$ GeV. 
Additionally, as seen in the same displays, 
significantly larger $m_0$ values than what are mentioned above are 
excluded for a limited region of $\m3half$.  

Constraining the minimal AMSB model via $\amususy$ 
can be further effective in the $(\smallchargino, \muonsneutrino)$ plane
(see Figs.(\ref{constraints_mpr}b) and (\ref{constraints_mpr}d))
because, as noted earlier, the chirality nondiagonal lighter 
chargino terms dominate over the other contributions.
A value of $\muonsneutrino$ below 225 (325) GeV for $\tanbeta=25 (40)$
is explicitly ruled out via the current limit on $\amususy$. Here we
note that in situations similar to the minimal AMSB model, where
$\smallchargino$ and $\smallneut$ masses are almost degenerate 
and sneutrinos are light, the present experimental lower bound of 
$\smallchargino$ is 56~GeV\cite{maltoni}.
The white regions denoted by {\it X} in the bottom of the $\amususy$ allowed 
and disallowed zones of Figs.(\ref{constraints_mpr}b) and 
(\ref{constraints_mpr}d) are disallowed for the same reasons
as mentioned before and we will further comment on them in Sec. 5 
while discussing similar regions in Figs.(\ref{constraints_mun}) and 
(\ref{constraints_pun}). 

\subsection{Probing the minimal AMSB scenario further via $\amususy$ and 
the BNL experiment at its predicted level of accuracy}
   The uncertainty level of $ {\delta \amu}^{BNL}=4 \times 10^{-10}$,
which is going to be achieved at Brookhaven within a few years, will, 
at least, significantly constrain the parameter space of a theory beyond the 
Standard Model.
Considering this predicted level of accuracy, we constrain 
$\amususy$ within the $2\sigma$ limit (see Figs.(\ref{constraints_mun}) and 
(\ref{constraints_pun})), 
where $\sigma$ is obtained from the 
predicted uncertainty level of BNL experiment and the error 
associated with the SM result, added in quadratures. The assumed nondiffering central estimates
of the experimental and theoretical results would be the limiting scenario 
of seeing no
deviation from the Standard Model result. 
This analysis would be valid in the situation when the experiment is 
complete and no deviation from the Standard Model is seen within the 
error limits.
This is in a similar line to what have been seen in Refs.\cite{ucpn,SUSYBNL} 
for supersymmetric as well as nonsupersymmetric theories.
On a further note, we assume that the hadronic error in $\amu^{had1}$
would be staying at its present level. A reduction, on the other hand, 
which will occur in the near future, would further constrain a 
similar analysis.

The lower triangular regions in the ($\m3half, m_0$) plane of Figs.
(\ref{constraints_mun}) and (\ref{constraints_pun}) 
are the disallowed zones, where 
$|\amususy|$ exceeds the 2$\sigma$ limit of the combined uncertainty.
The same result within 
the $\amususy$-relevant mass pairs ($\smallchargino, \muonsneutrino$)
is presented in the right hand sides of Figs.(\ref{constraints_mun}) and 
(\ref{constraints_pun}).  We note that, corresponding to 
$\tanbeta$=10 and 25, the minimal AMSB satisfied parameter space
is reasonably identical, with respect to the sign of $\mu$.
In Figs.(\ref{constraints_mun}e) and (\ref{constraints_pun}e), however, the 
regions differ in this aspect, and we will come back to them in Sec. 5.

Fig.(\ref{constraints_mun}) and (\ref{constraints_pun}) indicate that 
$m_0<$ 275 (475) GeV domains will be entirely  eliminated for 
$\tanbeta=$10 (25) for both signs of $\mu$ . For $\tanbeta=40$,
the corresponding limits are 625 GeV for $\mu<0$ and 800 GeV  
for $\mu>0$. The limit of $m_0 \leq 800$~GeV for $\mu>0$ appears because of
reasons which we will discuss soon.
Within the ($\smallchargino, \muonsneutrino$) planes of the right hand side
of Figs.(\ref{constraints_mun}) and (\ref{constraints_pun}), we find that 
for $\mu < 0$ and $\tanbeta=$ 10, 25, and 40,
values of $\muonsneutrino$ less than 210, 400, and 560 GeV are excluded.
The situation for $\mu>0$ is identical for $\tanbeta=$ 10 and 25.
A significant difference between the signs of $\mu$ can be
seen now, switching to $\mu>0$ and $\tanbeta=$40.
A similar disallowed range will be very stringent here; namely, 
$\muonsneutrino$ below 780 GeV will be excluded.

Interestingly, an important result is found,
when this analysis of $\amususy$ is combined with the 
$b \rightarrow s + \gamma$ constraint.
The constraint from $b \rightarrow s + \gamma$ within the minimal AMSB scenario
as analyzed in Ref.~\cite{feng_moroi_g-2} is somewhat complementary to what 
we find here from $\amususy$. This is because the $b \rightarrow s + \gamma$ 
calculation, which has many special features in AMSB models, puts severe 
mass limits for $\mu>0$ and much smaller limits for $\mu<0$. 
On the other hand, within the above scenario of 
seeing no deviation from the SM result once the experiment is performed
at the predicted level of accuracy, $\amususy$ limits in the minimal AMSB 
model impose a very significant constraint for both $\mu<0$ and 
$\mu>0$ cases.

\section{Generally disallowed parameter zones}
A discussion about the generally eliminated parameter space 
may be useful in studying the supersymmetric contribution to muon $(g-2)$ 
in a given SUSY model, because a combined constraint from 
$\amususy$, as well as from any generic disallowedness, results in  
simpler and definite predictions. Restricting the stau from becoming tachyonic 
corresponds to a significant constraint in AMSB models. 
In this section, by {\it allowedness} we mean
valid input parameters from the model, 
in addition to satisfying various experimental
lower bounds of sparticle masses, without reference to 
any $\amususy$ constraint.

\begin{figure}[hbt]
\centerline{\epsfig{file= 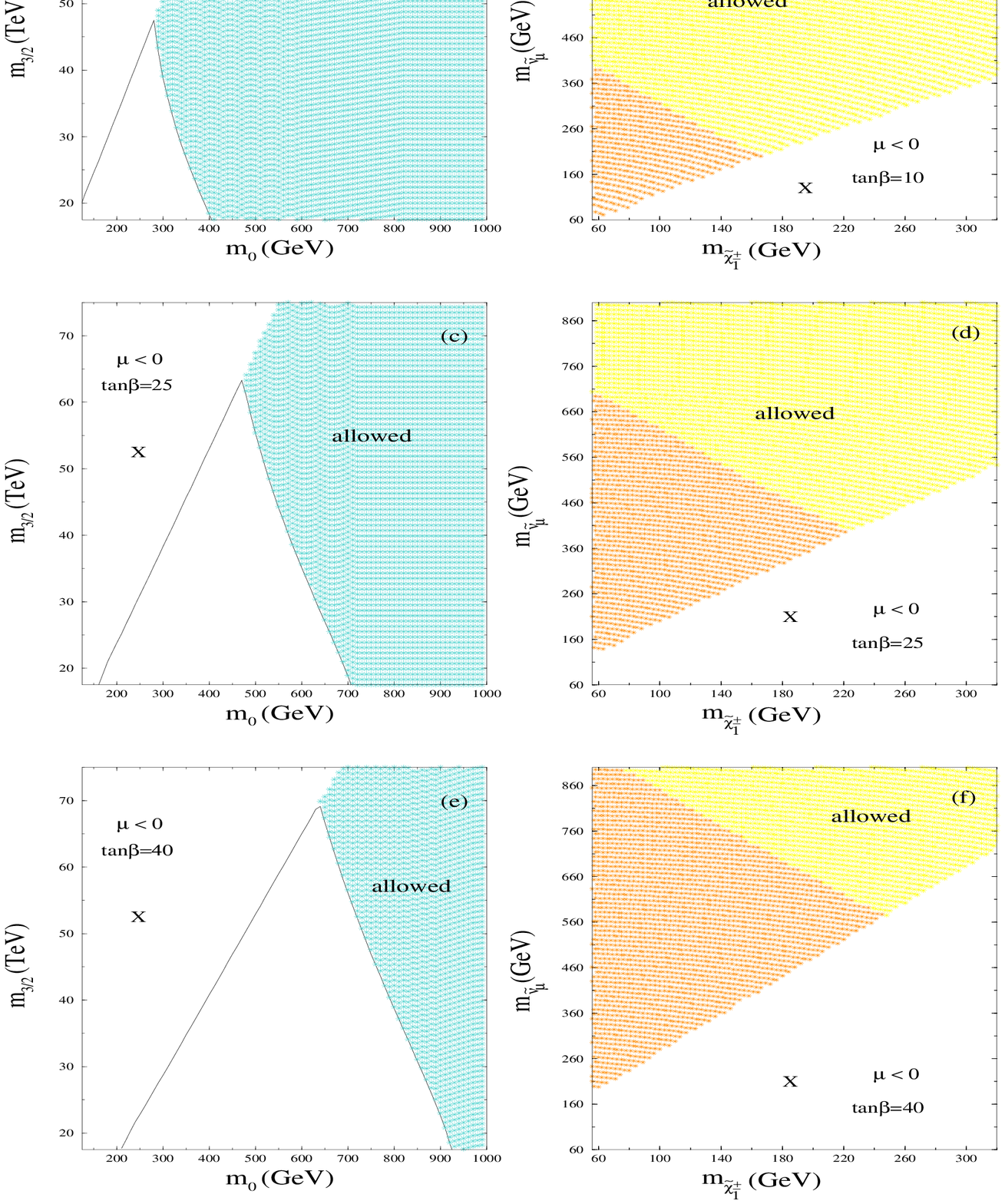,width=15cm}}
\vspace*{-2.0in}
\caption{Constraints in the $(m_0 - m_{3/2})$ and $(m_{\tilde \chi^\pm_1}- 
m_{\tilde \nu_\mu})$ planes within the {\it no deviation from SM} scenario
of $a_\mu^{\rm SUSY}$ as explained in the text for $\mu < 0$. 
Regions allowed 
by the $a_\mu^{\rm SUSY}$ constraint are labeled as {\it allowed}. Regions 
marked with {\it X} are generally disallowed zones for the SUSY spectra 
within minimal AMSB. White triangular regions below the {\it allowed} areas 
in the left hand side 
figures are disallowed via  the $a_\mu^{\rm SUSY}$ limit. The same regions in the 
right hand side are darkly shaded.
}
\label{constraints_mun}
\end{figure}

\begin{figure}[hbt]
\centerline{\epsfig{file= 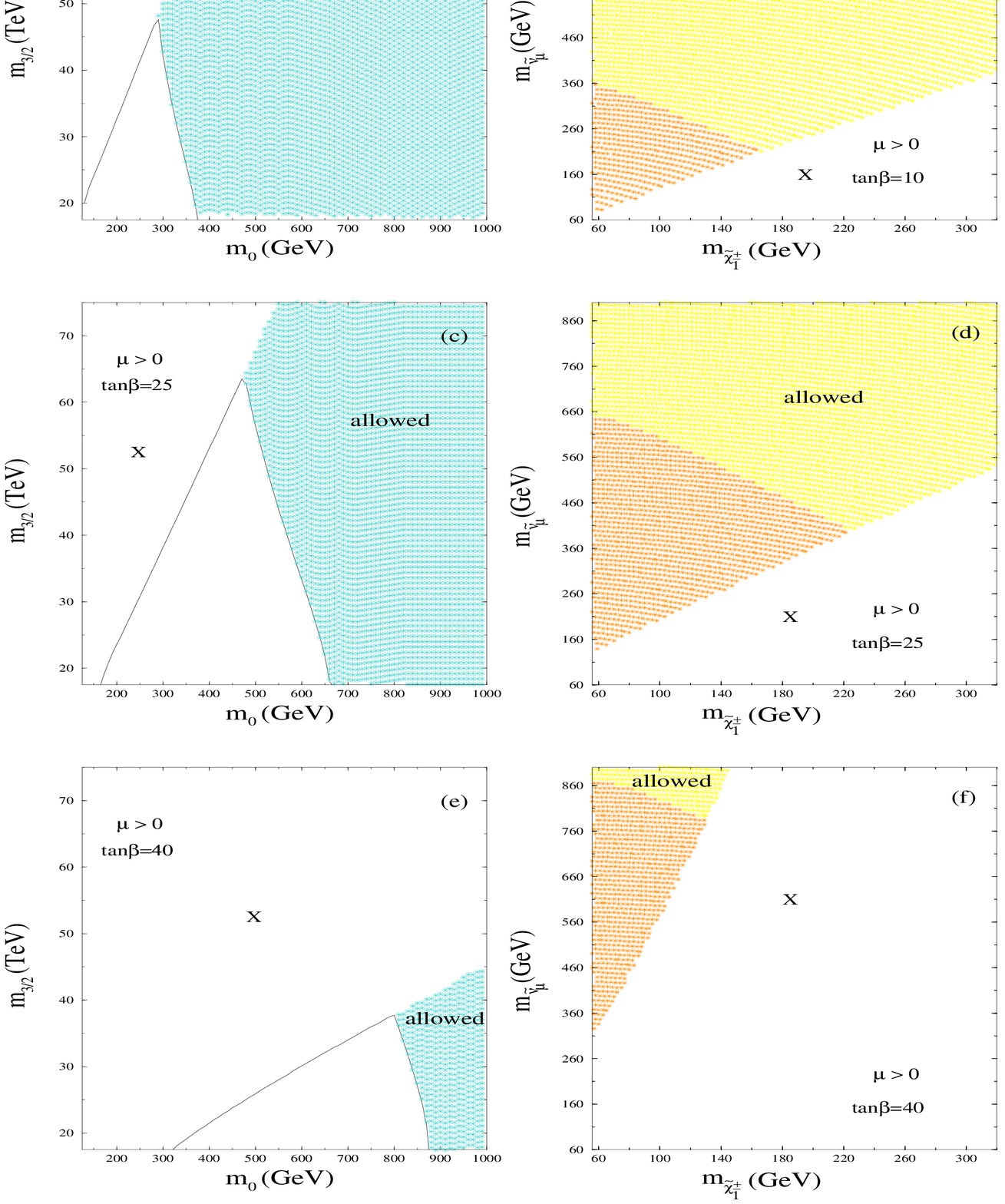,width=15cm}}
\vspace*{-2.0in}
\caption{Constraints the in ($m_0 - m_{3/2})$ and $(m_{\tilde \chi^\pm_1}- 
m_{\tilde \nu_\mu})$ planes within the {\it no deviation from SM} scenario
of $a_\mu^{\rm SUSY}$, as explained in the text for $ \mu > 0$.
Regions allowed 
by the $a_\mu^{\rm SUSY}$ constraint are labeled as {\it allowed}. Regions 
marked with {\it X} are generally disallowed zones for the SUSY spectra 
within minimal AMSB. White triangular regions below the {\it allowed} areas 
in the left hand side 
figures are disallowed via  the $a_\mu^{\rm SUSY}$ limit. The same regions in the 
right hand side are darkly shaded. 
}
\label{constraints_pun}
\end{figure}

We will first describe a few observations as revealed from our numerical 
analysis.
For a given $m_0$, the larger $\m3half$ values falling within the region 
labeled by {\it X} in Figs.~(\ref{constraints_mun}) 
and (\ref{constraints_pun}) ( also in Fig.~(\ref{constraints_mpr}) ) 
are eliminated because of a decreasing stau mass ($\staumass$), which either
goes below the experimental lower limit of 70 GeV\cite{stau70} or 
becomes the LSP, hence discarded in our {\it R}-parity 
conserved scenario. 
Thereafter, with a further increase of $\m3half$, 
the stau becomes tachyonic. We also see that the maximum 
possible $\m3half$ for a given $m_0$, as allowed by the minimal 
AMSB model, is larger for a smaller $\tanbeta$. 
Thus, for $\mu<0$ and  $m_0=$~400 GeV, comparing 
Figs.~(\ref{constraints_mun}a), (\ref{constraints_mun}c), and 
(\ref{constraints_mun}e) we find that such maximum possible 
values of $\m3half$ are approximately 67, 52, and 41 TeV 
for $\tanbeta=$~10, 25, and 40, respectively.
On the other hand, for a given $\tanbeta$, an allowed $\m3half$ increases 
for an increase in $m_0$. 
Besides, smaller $\m3half$ regions below the origins of 
the displays are eliminated via the experimental 
constraint of $\smallchargino \gsim $~56GeV\cite{maltoni}.

We will try to explain, qualitatively, the behavior of stau mass with
variations of the basic parameters of the model.
The effect of $\tilde \tau_1$ becoming tachyonic, as described above, is best 
explained via ${\tilde m}_L^2$ [see Eq.~(\ref{thirdgeneqn}] for $ i \equiv L$) 
assuming smaller left-right slepton mixing for convenience. There 
are two effects in ${\tilde m}_L^2$ due to a 
change in $\tanbeta$, which may support or oppose each other. The first one 
arises from the scale invariant part of ${\tilde m}_L^2$ and the other 
one originates from RG evolution\cite{martin94} of the same.  

The Yukawa term in the scale invariant
part of ${\tilde m}_L^2$ in Eq.~(\ref{thirdgeneqn}) is intrinsically 
negative, itself being also gauge coupling dominated within the corresponding 
$\hat \beta$ function, for the range of $\tanbeta$ considered 
in this analysis. 
Hence, the value of ${\tilde m}_L^2$ at
$M_G$ decreases if $\tanbeta$ is larger. 
We consider here moderate values of $m_0$ for a simpler discussion.
Until $\tanbeta$ is in a smaller domain, 
so that $\tau$-Yukawa coupling ($h_\tau$) 
within the scalar mass RG equation\cite{martin94} may be neglected compared 
to the gauge terms, the RGE effect due to running from $M_G$ 
to the electroweak scale always increases ${\tilde m}_L^2$ because of gauge 
domination.
Thus the two effects oppose each other. 
But within smaller $\tanbeta$ domains, 
regarding the value of ${\tilde m}_L^2$ at the electroweak 
scale for an increase in 
$\tanbeta$, the effect of the AMSB 
specified decrease in ${\tilde m}_L^2$ at $M_G$ 
is stronger than the increase due to the RGE effect. 
We have also verified this numerically in a broad domain of
parameter ranges.
As a result, ${\tilde m}_L^2$ and consequently  $\staumass$ decrease with 
an increase in $\tanbeta$. 
This in turn means that, for a given $m_0$, 
the upper limit of $\m3half$ is reached sooner for a 
larger $\tanbeta$.  This we may see from 
Figs.~(\ref{constraints_mun}) and (\ref{constraints_pun}), 
as well as from the values quoted above within this section. 

For a further increase in $\tanbeta\quad(\sim 40$ in our analysis), 
instead of opposing, the two effects may go in the same direction, 
although with varying strengths, because the $\tau$-Yukawa term may
now start to dominate within the RG evolution. In fact,
this may also be true when $m_0$ is large, with 
$\tanbeta$ in a moderately larger domain ($\sim$ 25).

On the other hand, corresponding to the lowest $\m3half$ values satisfying 
the lighter chargino experimental bound, a gradual increase of 
the lower limit of $m_0$ with an increase in $\tanbeta$ is found
(see left side displays of Fig.(\ref{constraints_mun}) and (\ref{constraints_pun})),
because, as explained before, the scale invariant part of 
${\tilde m}_L^2$ turns further negative for increasing $\tanbeta$, and
larger $m_0$ values are hence needed to compensate. However,
there is a marked difference between the $\mu<0$ and $\mu>0$ cases 
for $\tanbeta=40$ [see Fig.(\ref{constraints_mun}e) and (\ref{constraints_pun}e)]. 
The upper limit of the $\m3half$ for $\mu>0$ 
as allowed by the model, is much smaller compared to the same for $\mu<0$. This 
happens due to a generic large $\tanbeta$ effect, 
the effect of large SUSY-QCD loop corrections of bottom 
quark mass\cite{susybmass}.
Significantly, this correction has special features in the AMSB scenario~\cite{kribs}, 
because within the same the $SU(3)$ gaugino mass $\tilde m_3$ comes 
with a negative sign.
Consequently, for $\mu > 0$ and large $\tanbeta$, as a result of a large SUSY QCD loop
correction, a very large $h_b$ ($\sim h_t$) causes 
${m^2_{H_d}}$ for the Higgs scalar to turn sufficiently negative so 
that the CP-odd Higgs 
particle becomes tachyonic. However, here $\stau$ can still remain nontachyonic.
A further increase of 
$\m3half$ causes $\stau$ to become tachyonic, as usual.  

Considering now the 
combined effect of the model specified disallowed space, as well as the 
constraint from $\amususy$, we find that a region below $m_0=800$ GeV 
for $\mu>0$ 
will be completely eliminated within the scenario of seeing no deviation
from the SM. 
The right hand side displays of Figs.(\ref{constraints_mun}) and 
(\ref{constraints_pun}) also show model specified eliminated regions, as 
identified by {\it X} in the ($\smallchargino, \muonsneutrino$) plane. Obviously,
the masses are not independent, because 
they are derived from the basic set of input parameters of the minimal AMSB 
model.
The same reason for $\stau$ becoming 
tachyonic eliminates a large region within the zones {\it X}. 
The disallowed {\it X}-zone is large for $\mu>0$ and $\tanbeta=40$ in 
Fig.(\ref{constraints_pun}f), compared to the same for $\mu<0$, as shown in
Fig.(\ref{constraints_mun}f). This occurs because of the same reason for
which the upper limit of $\m3half$ is smaller in 
Fig.(\ref{constraints_pun}e), which we have explained before, 
and due to the fact that the lighter chargino is wino dominated 
within AMSB, thus $\smallchargino$ being almost proportional to $\m3half$. 

\section{Conclusion}
We have computed the supersymmetric contribution to the anomalous magnetic moment 
of the muon within the minimal Anomaly Mediated Supersymmetry Breaking model.
There are one-loop 
contributions involving chargino-sneutrino and neutralino-smuon parts.
The chiral interference term involving the lighter chargino is seen to contribute 
the most to $\amususy$ than the other chargino and neutralino terms, and this 
also results in a definite sign relationship between $\amususy$ and $\mu$. 
In addition, this also gives an almost proportional relationship of 
$|\amususy|$ with $\tanbeta$. 
We have also seen the effect of large smuon {\it L-R} mixing which causes
strong partial cancellations between the various terms of the neutralino result
of $\amususy$. This is a significantly important result of $\amususy$ within
minimal AMSB.

   We have analyzed the constraint coming from current values of
$\amu$ from the Standard Model and the ongoing experiment at Brookhaven,
assuming that the difference appears due to SUSY.
The constraint which exists only for $\mu < 0$ shows that, for $\tanbeta=$ 
25 (40), regions with $m_0 <$ 275 (375) GeV and, correspondingly, 
$\muonsneutrino<$ 225 (325) GeV are eliminated. We have also investigated the 
constraint from $\amususy$ that would result if the Brookhaven experiment 
with its already predicted level of accuracy finds no deviation from the 
Standard Model result. In this scenario, 
one finds that for $\mu < 0$  and $\tanbeta=$ 10, 25, and 40, the lower bounds
of $m_0$ would be 275, 475, and 625 GeV, while 
the corresponding lower limits for $\muonsneutrino$ would be 
210, 400, and 560 GeV, respectively. The lower bounds for $ \mu > 0$  are
identical to $\mu <$ 0, except for $\tanbeta=$40, where $m_0 < $~800 GeV 
and 
correspondingly $\muonsneutrino <$~780 GeV regions would be 
excluded. This happens due to a 
large SUSY-QCD correction to the bottom-quark mass, a large $\tanbeta$ effect.

  We have also compared our constraint with the same obtained
from $b \rightarrow s + \gamma$ from Ref.~\cite{feng_moroi_g-2}.
We found that the high accuracy level of the BNL experiment will
be very useful to constrain the model for $\mu < 0$, because the 
$b \rightarrow s + \gamma$ limit is effective only for $\mu >$ 0.  
Furthermore, we have also analyzed the generically disallowed 
zones within the parameter space of the model, because a combined 
constraint from $\amususy$ and such invalid parameter ranges 
lead to a stronger prediction. 

\vspace{0.5cm}
\noindent
{\large \bf{Acknowledgments:}}
DKG wishes to acknowledge the hospitality provided by the 
Theory Division, CERN, where part of this work was completed.

\newpage

\newpage
\begin{figure}[hbt]
\centerline{\epsfig{file=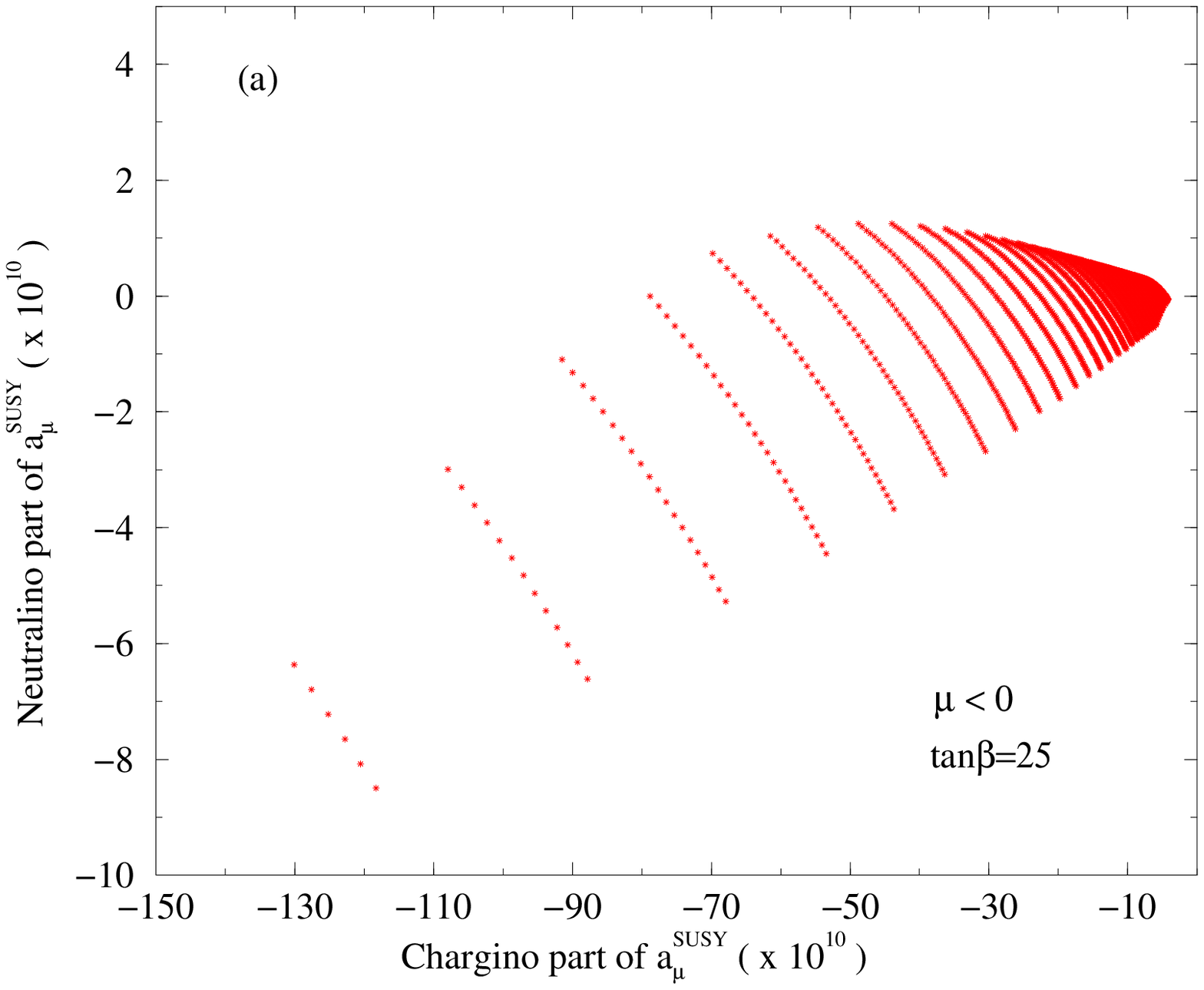,width=8cm}}
\end{figure}

\newpage

\newpage

\newpage

\end{document}